# A Gauge Model for Extra Weak Bosons Implied by Compositeness of Quarks and Leptons

Masaki Yasuè[*]

*Department of Natural Science, School of Marine Science and Technology*
*Tokai University, Shimizu, Shizuoka 424, Japan*

**Abstract**

Properties of extra weak bosons are discussed in a gauge model based on $SU(2)_L^{\ell oc} \times U(1)_Y^{\ell oc} \times \boldsymbol{\mathcal{SU}}(2)_L^{\ell oc}$. The physics of $\boldsymbol{\mathcal{SU}}(2)_L^{\ell oc}$ respects the duality in the Higgs and confining phases, where massive gauge bosons in the Higgs phase are equivalently represented by composite vector bosons in the confining phase below the scale of $\boldsymbol{\mathcal{SU}}(2)_L^{\ell oc}$. The interactions for $\boldsymbol{\mathcal{SU}}(2)_L^{\ell oc}$- singlet composites induced in the confining phase can be generated by nonlinear interactions of the four Fermi type respecting $SU(2)_L^{\ell oc} \times U(1)_Y^{\ell oc}$, which create quarks, $q_i^A$ (A=r,g,b; i=1,2), leptons, $\ell_i$, vector bosons, $V_\mu^{(a)}$ (a=1,2,3) and the Higgs scalar, $\phi$, made by scalar subconstituents, $c^{0,A}$, and spinor subconstituents, $w_i$, as $q_i^A \approx c^A w_i$, $\ell_i \approx c^0 w_i$, $V_\mu^{(a)} \approx \overline{w}\tau^{(a)}\gamma_\mu w$ and $\phi \approx \overline{w} w$. The physics described by these composites is equivalent to the one by $SU(2)_L^{\ell oc} \times U(1)_Y^{\ell oc} \times \boldsymbol{\mathcal{SU}}(2)_L^{\ell oc}$ below the confining scale.

---

[*] yasue@keyaki.cc.u-tokai.ac.jp

*Prologue*

Although extensive experimental studies [1] on Z properties at $e^+e^-$ colliders have almost confirmed the standard model of quarks and leptons based on the gauge theory of $SU(3)_c^{loc} \times SU(2)_L^{loc} \times U(1)_Y^{loc}$, there appear new experimental data indicating physics beyond the standard model, which include the excess of jets at large transverse energies at the $\bar{p}p$ collider [2] and of the events with large squared four momentum transfer at the $e^+p$ collider [3]. It is understood that any physics beyond the standard model must explain the origin of the scale of order of $G_F^{-1/2}$ governing weak interactions since $G_F^{-1/2}$ is not provided by the standard model. There are two main streams to generate $G_F^{-1/2}$: One is to introduce supersymmetry (SUSY) [4-6] that provides beautiful theoretical framework to manipulate radiative corrections, which in turn ascribe $G_F^{-1/2}$ to the scale of SUSY breaking and the other is to assume possible sizes of the order of $G_F^{1/2}$ for some of fundamental particles [7-9]. Lots of attempts have been made to explain new events by assuming SUSY or compositeness [10, 11].

Since SUSY models essentially rely upon the Higgs phases of assumed nonabelian gauge theories, their dynamical structure is well understood except for the SUSY breaking dynamics. On the other hand, dynamics of generating composite quarks and leptons has not yet clearly been understood. Recent theoretical understanding of N=1 SUSY confining gauge theories [12] greatly helps us toward constructing realistic composite models of quarks and leptons [13, 14]. The key features are duality and holomorphy, where the duality has originated from the study of ordinary gauge theories [15] without SUSY.

In the present article, we utilize the duality in confining gauge theories to describe compositeness of quarks and leptons and compare their interactions with those generated by another dynamical model based on nonlinear interactions of the four Fermi type to create composite quarks and leptons as well as Higgs scalar [8]. Our gauge model is based on $SU(2)_L^{loc} \times U(1)_Y^{loc} \times \boldsymbol{\mathcal{SU}}(2)_L^{loc}$ [16, 17], where $\boldsymbol{\mathcal{SU}}(2)_L^{loc}$ operates on left-handed states and gets confined to create composite quarks and leptons. The physics for composite quarks and leptons in the confining phase turns out to be the same as the one in the Higgs phase at "low" energies below the scale of $\boldsymbol{\mathcal{SU}}(2)_L^{loc}$. In the Higgs phase, the usual $SU(2)_L^{loc} \times U(1)_Y^{loc} \times \boldsymbol{\mathcal{SU}}(2)_L^{loc}$ gauge model shows up and gauge bosons are mixed with *W* and *Z* by mass terms, while in the confining phase, all particles including quarks and leptons are $\boldsymbol{\mathcal{SU}}(2)_L^{loc}$-singlets and composite vector bosons are mixed with *W* and *Z* by kinetic terms. We then demonstrate that the physics in the confining phase can be reproduced by nonlinear interactions of the four Fermi type among subconstituents, where the $SU(2)_L^{loc} \times U(1)_Y^{loc}$ symmetry is respected. Explicit evaluations on effective interactions for composites lead to precisely the same "low" energy interactions as those obtained in the confining phase, thereby, in the Higgs phase.

We discuss compositeness of quarks and leptons in $SU(2)_L^{loc} \times U(1)_Y^{loc} \times \mathcal{SU}(2)_L^{loc}$ and in its preonic version based on $SU(2)_L^{loc} \times U(1)_Y^{loc}$. Quantum numbers of quarks and leptons in $SU(2)_L^{loc} \times U(1)_Y^{loc} \times \mathcal{SU}(2)_L^{loc}$ are listed in **Table 1** as well as the Higgs scalar, $H$, and the additional scalar, $\xi$. To supply compositeness, the relevant fields must carry the $\mathcal{SU}(2)_L^{loc}$-color and do not carry the usual $SU(2)_L^{loc}$ quantum number. Composite quarks, leptons and Higgs scalar are expressed as:

$$u_L^A \approx \xi_1 q_L^A, \quad d_L^A \approx \xi_2 q_L^A, \quad \nu_{eL} \approx \xi_1 \ell_L, \quad e_L \approx \xi_2 \ell_L, \quad \phi \approx \xi H, \tag{1}$$

which are all $\mathcal{SU}(2)_L^{loc}$-singlets. In the preonic version, the corresponding compositeness is supplied by subconstituents, $w$ as spinors carrying weak isospin and $c$ as scalars carrying three colors and lepton number [7]:

$$u^A \approx c^A w_1, \quad d^A \approx c^A w_2, \quad \nu_e \approx c^0 w_1, \quad e \approx c^0 w_2, \quad \phi \approx \bar{w} w, \tag{2}$$

whose quantum numbers are listed in **Table 2**. Since the healthy gauge theory must not contain anomalies, the electric charges of $w$ is fixed to be (1/2, -1/2) for ($w_1$, $w_2$).

*Compositeness due to the duality*

We discuss the compositeness in the gauge model of $SU(2)_L^{loc} \times U(1)_Y^{loc} \times \mathcal{SU}(2)_L^{loc}$, where left-handed quarks and leptons and the Higgs scalar are taken to be doublets of $\mathcal{SU}(2)_L^{loc}$ but singlets of $SU(2)_L^{loc}$. The starting lagrangian is

$$\mathcal{L} = -\frac{1}{2}\mathrm{Tr}\left(W_{\mu\nu}^0 W^{0\mu\nu} + \mathcal{w}_{\mu\nu}\mathcal{w}^{\mu\nu}\right) - \frac{1}{4}B_{\mu\nu}B^{\mu\nu}$$
$$+ i\bar{q}_L\gamma_\mu\left[\partial^\mu - ig_\mathcal{w}\mathcal{w}^\mu - ig'\frac{Y}{2}B_\mu\right]q_L + i\bar{q}_R\gamma_\mu\left[\partial^\mu - ig'\frac{Y}{2}B_\mu\right]q_R$$
$$+ \bar{q}\left(H_q^\dagger L + H_q R\right)q + (q \to \ell) + \left|\left[\partial_\mu - ig_\mathcal{w}\mathcal{w}_\mu - ig'\frac{Y}{2}B_\mu\right]H\right|^2$$
$$+ \left|\left(\partial_\mu - igW_\mu^0\right)\xi + ig_\mathcal{w}\xi\mathcal{w}_\mu\right|^2, \tag{3}$$

with the Higgs potential, $V(H,\xi)$, whose explicit form is irrelevant for our present concern. The vector fields denoted by $W_\mu^0$, $B_\mu$ and $\mathcal{w}_\mu$ are gauge fields of $SU(2)_L^{loc} \times U(1)_Y^{loc} \times \mathcal{SU}(2)_L^{loc}$, $W_{\mu\nu}^0$, $\mathcal{w}_{\mu\nu}$ and $B_{\mu\nu}$ are the corresponding field strengths and $g$, $g'$ and $g_\mathcal{w}$ are their gauge couplings. The Higgs couplings are parametrized through $H_{q,\ell}$ defined by $H_q = (f_u H^G, f_d H)$ and $H_\ell = (f_\nu H^G, f_e H)$ with $H = (H_1, H_2)^\mathrm{T}$, $H^G = (-H_2^*, H_1^*)^\mathrm{T}$, where $f$'s are coupling strengths.

In the Higgs phase defined by $\langle \xi_i^\alpha \rangle = \Lambda_\xi \delta_i^\alpha$, vector bosons are mixed to be

$$W_\mu^D = \sin\theta_V \mathcal{W}_\mu + \cos\theta_V W_\mu^0, \quad V_\mu = \cos\theta_V \mathcal{W}_\mu - \sin\theta_V W_\mu^0, \tag{4}$$

where $\sin\theta_V = g/\sqrt{g_\mathcal{W}^2 + g^2}$. The gauge group $SU(2)_L^{loc} \times \mathcal{SU}(2)_L^{loc}$ spontaneously breaks down to its diagonal subgroup denoted by $SU(2)_D^{loc}$, where $W_\mu^D$ becomes its gauge boson whose gauge coupling is given by $g_D = g_\mathcal{W} \sin\theta_V = g\cos\theta_V$. The massive gauge boson, $V_\mu$, acquires a mass, $\sqrt{g^2 + g_\mathcal{W}^2}\Lambda_\xi$. By ignoring the Higgs potential and the physical fields from $\xi$, one finds $\mathcal{L}_{\text{Higgs}}$:

$$\begin{aligned}\mathcal{L}_{\text{Higgs}} = &-\frac{1}{2}\text{Tr}\{W_{\mu\nu}^D W^{D\mu\nu} + V_{\mu\nu}V^{\mu\nu} - 2g_D W_{\mu\nu}^D(i[V^\mu, V^\nu]) - 2\lambda_{VVV}^0 V_{\mu\nu}(i[V^\mu, V^\nu])\\ &+ \lambda_{VVVV}^0(i[V_\mu, V_\nu])(i[V^\mu, V^\nu])\} + m_V^2 Tr(V_\mu V^\mu) - \frac{1}{4}B_{\mu\nu}B^{\mu\nu}\\ &+ i\overline{q_L}\gamma^\mu\left[\partial_\mu - ig_D W_\mu^D - ig_\mathcal{W} c V_\mu - ig'\frac{Y}{2}B_\mu\right]q_L + i\overline{q_R}\gamma^\mu\left[\partial_\mu - ig'\frac{Y}{2}B_\mu\right]q_R\\ &+ \overline{q}(H_q^\dagger L + H_q R)q + (q\to\ell) + \left|\left[\partial_\mu - ig_D W_\mu^D - ig_\mathcal{W} c V_\mu - ig'\frac{Y}{2}B_\mu\right]H\right|^2,\end{aligned} \tag{5}$$

where

$$\begin{aligned}&m_V = \sqrt{g_\mathcal{W}^2 + g^2}\Lambda_\xi, \quad \lambda_{VVV}^0 = g_\mathcal{W} c^3 - gs^3, \quad \lambda_{VVVV}^0 = g^2 s^4 + g_\mathcal{W}^2 c^4,\\ &V_{\mu\nu}^0 = \partial_\mu V_\nu^0 - \partial_\nu V_\mu^0 - ig_D([W_\mu^D, V_\nu] - [W_\nu^D, V_\mu]),\end{aligned} \tag{6}$$

for $c = \cos\theta_V$ and $s = \sin\theta_V$.

Let us next proceed to examine the confining phase of $\mathcal{SU}(2)_L^{loc}$. Physical particles are all $\mathcal{SU}(2)_L^{loc}$-singlet composite fields defined as

$$q_{iL}^A = \xi_i^\alpha q_{\alpha L}^A/\Lambda_\xi, \quad l_{iL} = \xi_i^\alpha \ell_{\alpha L}/\Lambda_\xi, \quad ig_V V_\mu^0 = \frac{\xi^\dagger(\partial_\mu \xi + ig_\mathcal{W} \xi \mathcal{W}_\mu)}{\Lambda_\xi^2} - igW_\mu^0,$$

$$\phi_i = \xi_i^\alpha H_\alpha/\Lambda_\xi. \tag{7}$$

The confining phase is assumed to be characterized by

$$\left(\xi^\dagger\right)_i^\alpha \xi_\alpha^j = \Lambda_\xi^2 \delta_i^j, \quad \xi_\alpha^i\left(\xi^\dagger\right)_i^\beta = \Lambda_\xi^2 \delta_\alpha^\beta. \tag{8}$$

Inserting these composites in the lagrangian, (3), leads to $\mathcal{L}_{\text{conf}}$ calculated to be:

$$\begin{aligned}\mathcal{L}_{\text{conf}} = &-\frac{1}{2}\text{Tr} W_{\mu\nu}^0 W^{0\mu\nu} - \frac{1}{2g_\mathcal{W}^2}\text{Tr}(gW_{\mu\nu}^0 + g_V V_{\mu\nu}^0 + ig_V^2[V_\mu^0, V_\nu^0])^2 + g_V^2 \Lambda_\xi^2 \text{Tr}(V_\mu^0 V^{0\mu})\\ &- \frac{1}{4}B_{\mu\nu}B^{\mu\nu} + i\overline{q_L}\gamma_\mu\left[\partial^\mu - igW_\mu^0 - ig_V V_\mu^0 - ig'\frac{Y}{2}B_\mu\right]q_L\\ &+ i\overline{q_R}\gamma_\mu\left[\partial^\mu - ig'\frac{Y}{2}B_\mu\right]q_R - (\overline{q_R}\Phi_q^\dagger q_L + \overline{q_L}\Phi_q q_R) + (q\to\ell)\\ &+ \left|\left[\partial_\mu - igW_\mu^0 - ig_V V_\mu^0 - ig'\frac{Y}{2}B_\mu\right]\phi\right|^2,\end{aligned} \tag{9}$$

where

$$\tag{10}$$

$$V^0_{\mu\nu} = \partial_\mu V^0_\nu - \partial_\nu V^0_\mu - ig([W^0_\mu, V^0_\nu] - [W^0_\nu, V^0_\mu]).$$

To get the canonical kinetic terms, $W^0_\mu$ and $V^0_\mu$ are rescaled to $W^R_\mu$ and $V^R_\mu$ as

$$W^0_\mu = \frac{g_w}{\sqrt{g_w^2 + g^2}} W^R_\mu, \qquad V^0_\mu = \frac{g_w}{g_V} V^R_\mu, \tag{11}$$

yielding

$$\begin{aligned}\mathcal{L}_{\text{conf}} =& -\frac{1}{2}\text{Tr}\Big\{W^R_{\mu\nu}W^{R\mu\nu} + V^R_{\mu\nu}V^{R\mu\nu} + 2\lambda W^R_{\mu\nu}V^{R\mu\nu} - 2(g_D W^R_{\mu\nu} + g_w V^R_{\mu\nu})(i[V^{R\mu}, V^{R\nu}]) \\ &+ g_w^2(i[V^R_\mu, V^R_\nu])(i[V^{R\mu}, V^{R\nu}])\Big\} + g_w^2 \Lambda_\xi^2 \text{Tr}(V^R_\mu V^{R\mu}) - \frac{1}{4}B_{\mu\nu}B^{\mu\nu} \\ &+ i\overline{q}_L \gamma_\mu\left[\partial^\mu - ig_D W^R_\mu - ig_w V^R_\mu - ig'\frac{Y}{2}B_\mu\right]q_L + i\overline{q}_R\gamma_\mu\left[\partial^\mu - ig'\frac{Y}{2}B_\mu\right]q_R \\ &-(\overline{q}_R\Phi_q^\dagger q_L + \overline{q}_L\Phi_q q_R) + (q\to\ell) + \left|\left[\partial_\mu - ig_D W^R_\mu - ig_w V^R_\mu - ig'\frac{Y}{2}B_\mu\right]\phi\right|^2.\end{aligned} \tag{12}$$

where $\lambda = g_D/g_w = \sin\theta_V$ that controls the kinetic mixing of the gauge field and composite vector field. The last task is, thus, to remove the kinetic mixing, which is achieved by introducing physical fields, $W^D_\mu$ and $V_\mu$:

$$W^D_\mu = W^R_\mu + \lambda V^R_\mu, \qquad V_\mu = \sqrt{1-\lambda^2}\, V^R_\mu. \tag{13}$$

The resulting lagrangian is found to be:

$$\begin{aligned}\mathcal{L}_{\text{conf}} =& -\frac{1}{2}\text{Tr}\Big\{W^D_{\mu\nu}W^{D\mu\nu} + V_{\mu\nu}V^{\mu\nu} - 2g_D W^D_{\mu\nu}(i[V^\mu, V^\nu]) - 2\lambda_{VVV}V_{\mu\nu}(i[V^\mu, V^\nu]) \\ &+ \lambda_{VVVV}(i[V^\mu, V^\nu])(i[V^\mu, V^\nu])\Big\} + \mu_V^2 \text{Tr}(V_\mu V^\mu) - \frac{1}{4}B_{\mu\nu}B^{\mu\nu} \\ &+ i\overline{q}_L\gamma_\mu\left[\partial^\mu - ig_D W^D_\mu - ig^* V_\mu - ig'\frac{Y}{2}B_\mu\right]q_L + i\overline{q}_R\gamma_\mu\left[\partial^\mu - ig'\frac{Y}{2}B_\mu\right]q_R \\ &-(\overline{q}_R\Phi_q^\dagger q_L + \overline{q}_L\Phi_q q_R) + (q\to\ell) + \left|\left[\partial_\mu - ig_D W^D_\mu - ig^* V_\mu - ig'\frac{Y}{2}B_\mu\right]\phi\right|^2.\end{aligned} \tag{14}$$

where

$$\begin{aligned}\mu_V &= g_w \Lambda_\xi / \sqrt{1-\lambda^2}, \quad g^* = (g_w - \lambda g_D)/\sqrt{1-\lambda^2}, \\ \lambda_{VVV} &= \left[g_w + g_D \lambda(2\lambda^2 - 3)\right]/(1-\lambda^2)^{2/3}, \\ \lambda_{VVVV} &= \left[g_w^2 - 4g_D g_w \lambda - 3g_D^2 \lambda^2(\lambda^2 - 2)\right]/(1-\lambda^2)^2.\end{aligned} \tag{15}$$

The equivalence of $\mathcal{L}_{\text{conf}}$, (14), to $\mathcal{L}_{\text{Higgs}}$, (5), is proved by exhibiting $\mu_V = m_V$, $g^* = \cos\theta_V g_w$, $\lambda_{VVV} = \lambda^0_{VVV}$ and $\lambda_{VVVV} = \lambda^0_{VVVV}$ with the identification of $q_L = q_L$ and $\phi = H$. Since $\lambda = g_D/g_w = \sin\theta_V$, it is straightforward to show these equalities. Thus, one can find that the "low" energy lagrangian of $SU(2)^{\ell oc}_L \times U(1)^{\ell oc}_Y \times \mathcal{SU}(2)^{\ell oc}_L$ is the same for the elementary particles and for the composite particles.

*Compositeness due to nonlinear interactions of the four Fermi type*

To realize the compositeness in the preonic version, we adopt nonlinear interactions of the four Fermi-type. Composite quarks, leptons and the Higgs scalar are assumed to be generated by the following lagrangian, $\mathcal{L}_{int}$ [16]:

$$\mathcal{L}_{int} = \frac{1}{\Lambda_q}\left[i\bar{f}(D_\mu c)\gamma^\mu w - i\bar{w}(D_\mu c)^\dagger \gamma^\mu f\right] - \bar{f}(\Phi_f^\dagger L + \Phi_f R)f$$
$$- \bar{w}(\Phi_w^\dagger L + \Phi_w R)w - \Lambda_\phi^2 |\phi|^2 \tag{16}$$

where $\Phi_w = (h_{w_1}\phi^G, h_{w_2}\phi)$ and $\Phi_f = (h_u\phi^G, h_d\phi)$ or $= (h_\nu\phi^G, h_\ell\phi)$ with $\phi = (\phi_1, \phi_2)^T$ and $\phi^G = (-\phi_2^*, \phi_1^*)^T$, $h$'s are coupling constants, $D_\mu$ is the appropriate covariant derivative involving gauge bosons of $SU_c^{loc}(3) \times SU_L^{loc}(2) \times U(1)_Y^{loc}$ and $\Lambda_{q,\phi}$ represent the compositeness scale, $\Lambda_{comp}$, of the order of 1 TeV. The chirality conserving interactions are taken to create quarks and leptons. The compositeness of quarks ($f^A$), leptons ($f^0$) and Higgs scalar dictates from (16) as

$$f_i^{A,0} = i\frac{(\partial c^{A,0})\left(h_f^{-2}(\Phi_f^\dagger L + \Phi_f R)\right)_{ij}}{\Lambda_q |\phi|^2} w_j + (\cdots), \tag{17}$$

$$\phi = \frac{h_{w_2}\overline{w_L}^G w_{1R} - h_{w_1}\overline{w_{2R}} w_L}{\Lambda_\phi^2} + (\cdots) \tag{18}$$

with $h_f^{-2} = \text{diag}(h_u^{-2}, h_d^{-2})$ or $= \text{diag}(h_\nu^{-2}, h_\ell^{-2})$ and $\overline{w_L}^G = (-\overline{w_{L2}}, \overline{w_{L1}})^T$, which give our dynamical realization of the simple-minded compositeness (2). Substituting (17) and (18) into $\mathcal{L}_{int}$ finally yields the interaction lagrangian expressed in terms of the subconstituents only. The composite vector boson is created by four Fermi interactions of $w$:

$$\mathcal{L}_V = -\frac{1}{2\Lambda_V^2}\left(\overline{w_L}\gamma_\mu \frac{\tau^{(a)}}{2} w_L\right)^2, \tag{19}$$

where $\tau^{(a)}$ (a=1,2,3) is the weak isospin matrix and $\Lambda_V \sim \Lambda_{comp}$. The compositeness is explicitly given by translating the four Fermi interactions in terms of the auxiliary filed, $V_\mu^0$, into

$$\mathcal{L}_V = g_V \overline{w_L}\gamma^\mu V_\mu^0 w_L + \frac{\mu_V^{02}}{2}\left(V_\mu^{0(a)} V^{0(a)\mu}\right)^2, \tag{20}$$

where

$$V_\mu^{0(a)} = -\frac{1}{g_V \Lambda_V^2}\left(\overline{w_L}\gamma_\mu \frac{\tau^{(a)}}{2} w_L\right) \tag{21}$$

with $\mu_V^0 = g_V \Lambda_V$. Collecting these interactions, we are ready to construct our whole lagrangian, $\mathcal{L}_{tot} = \mathcal{L}_0 + \mathcal{L}_{int} + \mathcal{L}_V$ with

$$\mathcal{L}_0 = i\overline{w_L}\gamma^\mu \left(\partial_\mu - igW_\mu^0 - ig'\frac{Y}{2}B_\mu\right)w_L + i\overline{w_{iR}}\gamma^\mu\left(\partial_\mu - ig'\frac{Y_i}{2}B_\mu\right)w_{iR}$$

$$+ \left|\left(\partial_\mu - ig_c G_\mu - ig'\frac{Y_c}{2}B_\mu\right)c\right|^2 + \left|\left(\partial_\mu - ig'\frac{Y_0}{2}B_\mu\right)c^0\right|^2 - \left(m_c^2|c^A|^2 + m_0^2|c^0|^2\right)$$

$$-\frac{1}{2}Tr\left(G_{\mu\nu}G^{\mu\nu} + W_{\mu\nu}^0 W^{0\mu\nu}\right) - \frac{1}{4}B_{\mu\nu}B^{\mu\nu}, \qquad (22)$$

where $G_\mu$ ($g_c$) are gauge field (gauge couplings) of $SU(3)_c^{loc}$ and $m_{c,0}$ denote the masses of $c^{A,0}$.

Since below the scale, $\Lambda_{comp}$, subconstituents are all confined, effective interactions do not involve subconstituents. To realize such dynamical situation, one can construct the effective lagrangian, $\mathcal{L}_{eff}$, for composites by integrating out all subconstituents, which is defined by

$$\exp\left[i\int d^4x \mathcal{L}_{eff}\right] = \int [d\overline{w}][dw][dc^\dagger][dc]\exp\left[i\int d^4x \mathcal{L}_{tot}\right]. \qquad (23)$$

After performing the path integration, one finds, at the leading order, the compositeness conditions:

$$\frac{g_V^2 NJ_0}{3} = 1, \quad (f_{w_{11}}^2 + f_{w_2}^2)NJ_0 = 1, \quad \frac{5J_2}{4\Lambda_f^2} = 1, \qquad (24)$$

where $N$ denotes the number of copies of $w$ that is equal to the number of "colors" if our nonlinear interactions originate from a confined $SU(N)^{loc}$ force. These relations determine $g_V$, $f_{w_{1,2}}^2$ and $\Lambda_f$. The divergent integrals, $J_{0,2}$, are defined by:

$$J_{2n} = (-1)^{n+1}\int \frac{d^4k}{(2\pi)^4}\frac{1}{(k^2 - \overline{m}^2)^{2-n}}, \qquad (25)$$

for $n=0,2$, where $\overline{m}$ stands for average masses of $s$ and $b$. These integrals are regulated, by introducing a cutoff $\Lambda$ of the order of $\Lambda_{comp}$, to be $J_2 = \Lambda^2/(4\pi)^2$ and $J_0 = \ln(\Lambda/\overline{m})^2/(4\pi)^2$, where we have made a gauge invariant regularization with all insertions of the subconstituent's masses included in their propagators.

The explicit form of the evaluated lagrangian is found to be:

$$\mathcal{L}_{eff} = -\frac{1}{2}Tr\{W_{\mu\nu}^0 W^{0\mu\nu} + V_{\mu\nu}^0 V^{0\mu\nu} + 2\lambda^0 W_{\mu\nu}^0 V^{0\mu\nu} + 2(gW_{\mu\nu}^0 + g_V V_{\mu\nu}^0)(i[V^{0\mu},V^{0\nu}])$$

$$+ g_V^2(i[V_\mu^0,V_\nu^0])(i[V^{0\mu},V^{0\nu}])\} + \mu_V^{02}Tr(V_\mu^0 V^{0\mu}) - \frac{1}{4}B_{\mu\nu}B^{\mu\nu}$$

$$+ i\overline{q_L}\gamma_\mu\left[\partial^\mu - igW_\mu^0 - ig_V V_\mu^0 - ig'\frac{Y}{2}B_\mu\right]q_L + i\overline{q_R}\gamma_\mu\left[\partial^\mu - ig'\frac{Y}{2}B_\mu\right]q_R$$

$$-\bar{q}\left((\Phi_w + \Phi_q)^\dagger L + (\Phi_w + \Phi_q)R\right)q + (q \to \ell)$$

$$+ \left|\left[\partial_\mu - igW_\mu^0 - ig_V \boldsymbol{V}_\mu^0 - ig'\frac{Y}{2}B_\mu\right]\phi\right|^2 - \mu_\phi^2|\phi|^2 - \lambda_\phi|\phi|^4, \tag{26}$$

with appropriate $SU(3)_c^{loc}$ gauge interactions, where

$$\lambda^0 = g/g_V, \quad \lambda_\phi = \sum_{i=1,2} f_{w_i}^2 \Big/ \sum_{i=1,2} f_{w_i}^4, \quad \mu_\phi^2 = \Lambda_\phi^2 - (8\Lambda_f^2/5)(\sum_{i=1,2} f_{w_i}^2)N. \tag{27}$$

The spontaneous symmetry breaking of $SU(2)_L^{loc}$ is generated for $\mu_\phi^2 < 0$, which we require.

Now, we are ready to discuss the connection between the derived effective lagrangians for composites: one is based on the confining phase of $\mathcal{SU}(2)_L^{loc}$, (12), and the other on the nonlinear interactions, (26). By replacing fields and couplings according to

$$W_\mu^0 \Leftrightarrow W_\mu^R, \quad \boldsymbol{V}_\mu^0 \Leftrightarrow \boldsymbol{V}_\mu^R, \quad \Phi_w + \Phi_f \Leftrightarrow \Phi_f, \quad \boldsymbol{q}_R \Leftrightarrow q_R, \quad \boldsymbol{l}_R \Leftrightarrow \ell_R,$$
$$\mu_V^0 = g_V \Lambda_V \Leftrightarrow g_\mathcal{W} \Lambda_\xi, \quad g(g_V) \Leftrightarrow g_D(g_\mathcal{W}), \tag{28}$$

one finds that all couplings for quarks, leptons and vector bosons precisely coincide with each other. Therefore, only one "low"-energy physics appears to govern these composite particles.

*Effects of extra weak gauge bosons on weak interactions*

Effective "low" energy weak interactions are mediated by $W$ and $Z$ as well as extra weak bosons, $W'$ and $Z'$, which are composites or elementary. Their mass terms generated by the spontaneous breaking of $SU(2)_L^{loc} \times U(1)_Y^{loc}$ are given through the Higgs kinetic term:

$$\left|\left[\partial_\mu - igW_\mu^0 - ig_V \boldsymbol{V}_\mu^0 - ig'\frac{Y}{2}B_\mu\right]\phi\right|^2, \tag{29}$$

Let $\langle\phi\rangle = (0, v/\sqrt{2})^T$, then massless gauge bosons, $W_\mu^0$ and $B_\mu$, acquire masses and mix with the composite vector bosons, $\boldsymbol{V}_\mu^0$. These mixings are characterized by mass matrices, $M^{ch}$ and $M^n$:

$$W_\mu^D = W_\mu^0 + \lambda \boldsymbol{V}_\mu^0, \qquad V_\mu = \sqrt{1-\lambda^2}\,\boldsymbol{V}_\mu^0 \tag{30}$$

with their kinetic mixings removed. On the $(W_\mu^{D(\pm)}, V_\mu^{(\pm)})$-basis for $M^{ch}$ and on the $(Z_\mu^D, V_\mu^{(3)})$-basis for $M^n$ with $Z_\mu^D = \cos\theta W_\mu^{D(3)} - \sin\theta B_\mu$ and $\sin\theta = e/g$, the mixings are given

$$M^{ch} = \begin{bmatrix} m_{W0}^2 & \varepsilon m_{W0}^2 \\ \varepsilon m_{W0}^2 & \varepsilon^2 m_{W0}^2 + m_V^2 \end{bmatrix}, \quad M^n = \begin{bmatrix} m_{Z0}^2 & \varepsilon\cos\theta\, m_{Z0}^2 \\ \varepsilon\cos\theta\, m_{Z0}^2 & \varepsilon^2 m_{W0}^2 + m_V^2 \end{bmatrix}, \tag{31}$$

where
$$m_{W0} = \cos\theta m_{Z0} = gv/2, \qquad \varepsilon = [(g_V/g) - \lambda]/\sqrt{1-\lambda^2} = \sqrt{1-\lambda^2}/\lambda. \qquad (32)$$

The mass matrices are diagonalized to give $(m_W, m_{W'}, m_Z, m_{Z'})$ for physical fields, $(W_\mu^\pm, W_\mu'^\pm, Z_\mu, Z_\mu')$ defined by

$$\begin{pmatrix} W_\mu^\pm \\ W_\mu'^\pm \end{pmatrix} = \begin{pmatrix} c_\delta & -s_\delta \\ s_\delta & c_\delta \end{pmatrix} \begin{pmatrix} W_\mu^{D(\pm)} \\ V_\mu^{(\pm)} \end{pmatrix}, \quad \begin{pmatrix} Z_\mu \\ Z_\mu' \end{pmatrix} = \begin{pmatrix} c_\alpha & -s_\alpha \\ s_\alpha & c_\alpha \end{pmatrix} \begin{pmatrix} Z_\mu^D \\ V_\mu^{(3)} \end{pmatrix}, \qquad (33)$$

for $s_{\alpha,\delta} \geq 0$, where $c_\delta = \cos\delta$, $s_\delta = \sin\delta$, etc. By just looking at (31), one can find the useful mass relation [18]:

$$m_W m_{W'} = \cos\theta m_Z m_{Z'}, \qquad (34)$$

which is the generalization of $m_W = \cos\theta m_Z$ in the standard model. The mixing angles, $\alpha$ and $\delta$, can be expressed by these masses:

$$s_\delta^2 = \left(c_\theta^2 m_{Z'}^2 - m_W^2\right)\left(m_{Z'}^2 - m_{W'}^2\right)/s_\theta^2 m_{Z'}^2 \left(m_{W'}^2 - m_W^2\right), \qquad (35)$$

$$s_\alpha^2 = \left(m_{Z'}^2 - m_W^2\right)\left(m_{Z'}^2 - m_{W'}^2\right)/s_\theta^2 m_{Z'}^2 \left(m_{Z'}^2 - m_Z^2\right), \qquad (36)$$

which call for $m_{Z'} \geq m_{W'}$ that in turn yields

$$\cos\theta \leq \cos\theta_{WS} \text{ with } \cos\theta_{WS} = m_W/m_Z. \qquad (37)$$

It should be noted that mixings vanish at $m_{W'} = m_{Z'}$.

The contributions from $W'$ and $Z'$ to weak interactions are specified by

$$\mathcal{L}_{weak} = \frac{g}{2}\left[J_\mu^+\left(W^{(-)\mu} + \varepsilon V^{(-)\mu}\right) + (\text{H.C.})\right] + g_Z J_\mu^Z Z^\mu + g\varepsilon J_\mu^3 V_\mu^{(3)} + eJ_\mu A^\mu, \qquad (38)$$

where $J_\mu^Z = J_\mu^3 - s_\theta^2 J_\mu^{em}$ and $g_Z = g/\sqrt{g^2 + g'^2}$. Their couplings to quarks and leptons are parametrized as $V_W - A_W\gamma_5$ for $W$, $V_Z - A_Z\gamma_5$ for $Z$, $V_W' - A_W'\gamma_5$ for $W'$ and $V_Z' - A_Z'\gamma_5$ for $Z'$:

$$V_W = A_W = g(c_\delta - \varepsilon s_\delta),$$
$$V_W' = A_W' = g(\varepsilon c_\delta + s_\delta),$$
$$V_Z = g_Z\left(\eta I^{(3)} - 2c_\alpha s_\theta^2 Q_{em}\right) = g_Z\eta\left(I^{(3)} - 2\sin^2\theta_{eff} Q_{em}\right),$$
$$A_Z = g_Z\eta I^{(3)},$$
$$V_Z' = g_Z\left(\eta' I^{(3)} - 2s_\alpha s_\theta^2 Q_{em}\right) = g_Z\eta'\left(I^{(3)} - 2\sin^2\theta'_{eff} Q_{em}\right),$$
$$A_Z' = g_Z\eta' I^{(3)}, \qquad (39)$$

where

$$\eta = c_\alpha - \varepsilon s_\alpha c_\theta, \quad \eta' = \varepsilon c_\alpha c_\theta + s_\alpha,$$
$$\sin^2\theta_{eff} = c_\alpha \sin^2\theta/\eta, \quad \sin^2\theta'_{eff} = s_\alpha \sin^2\theta/\eta. \qquad (40)$$

To relate $G_F$ of weak interaction with the weak boson masses requires evaluating the low-energy limit of $\mathcal{L}_{weak}$, which is calculated to be $\mathcal{L}_{eff}^{ch}$ for the $W$ and $W'$ exchanges, and $\mathcal{L}_{eff}^n$ for the $Z$ and $Z'$ exchanges:

$$\mathcal{L}_{eff}^{ch} = 2\sqrt{2} G_F J^+ J^-, \tag{41}$$

$$\mathcal{L}_{eff}^{n} = 4\sqrt{2} G_F \left[ J^Z J^Z + \varepsilon_{em}^2 \sin^4\theta J^{em} J^{em} \right], \tag{42}$$

with the Lorentz indices are suppressed, where

$$4\sqrt{2} G_F = \rho g^2 / m_W^2, \quad \varepsilon_{em} = \varepsilon (m_W / m_{W'}), \tag{43}$$

for $\rho = (c_\delta - \varepsilon s_\delta)^2 + (s_\delta + \varepsilon c_\delta)^2 (m_W / m_{W'})^2$. Once $G_F$ is fixed, the low energy modification only participates in the parity-conserving neutral current interactions suppressed by the coupling strength squared of the extra bosons, $\varepsilon^2$.

*Epilogue*

We have shown two types of dynamical composite models of quarks, leptons and Higgs scalar, which involve extra weak bosons, $W'$ and $Z'$. The low-energy physics described by both models turn out to be the same as far as scalar excitations are frozen. The physical equivalence between massive gauge bosons and composite vector bosons has been advocated in earlier references [19]. Of course, the differences arise if we go to higher energies. In the model based on the confining $\mathcal{SU}(2)_L^{\ell oc}$, the $\mathcal{SU}(2)_L^{\ell oc}$-"gluons" emerge instead of $W'$ and $Z'$ while in the model based on preonic configurations, $W'$ and $Z'$ are disolved into the subconstituents, $w$ and $\bar{w}$. This situationn is shown in **Figure 1**. The duality in $\mathcal{SU}(2)_L^{\ell oc}$ ensures that effective interactions for composites are identical to the spontaneously broken $\mathcal{SU}(2)_L^{\ell oc}$, where all particles are elementary objects. Thus, if $W'$ and $Z'$ are observed as "elementary" particles, it does not directly suggest that they are massive gauge bosons associated with the broken $\mathcal{SU}(2)_L^{\ell oc}$. The present quantum number assignment of $\mathcal{SU}(2)_L^{\ell oc}$ for quarks and leptons are motivated by implementing their compositeness into the gauge model. Another interesting case is to require that $\mathcal{SU}(2)_L^{\ell oc}$ only operates on quarks. Then, quarks become composites and interacts with extra $\mathcal{SU}(2)_L^{\ell oc}$ bosons while leptons remain as "elementary" particles. This case corresponds to "leptophobic" [20]. The preliminary analysis on the "leptophobic" case has been performed in Ref.[21]. The other case, where all quarks and leptons are "elementary", is called "fermiophobic" [22].

In the present context, there is no reason why most of observed quarks and leptons are so light. One solution is to invoke the chiral $SU(4)$ version of the Pati-Salam symmetry [7] for $c^{A,0}$ with the appropriate anomaly-matching conditions satisfied [23]. To implement this symmetry calls for fermionic $c^{A,0}$ instead of the scalar $c^{A,0}$. Then, quarks and leptons become of the $whc$-type [8]. Another solution may utilize SUSY with Nambu-Goldstone supermultiplets on the coset space such as $SU(6)_L \times SU(6)_R / SU(4)_L \times SU(4)_R \times SU(2)_{L+R}$, whose fermionic components are of the $cw$-type [13]. With this possibility of the SUSY extension in mind, the analyses on the SUSY version of $SU(2)_L^{\ell oc} \times U(1)_Y^{\ell oc} \times \mathcal{SU}(2)_L^{\ell oc}$, which will corresponds to the preonic SUSY configuration of $SU(2)_L^{\ell oc} \times U(1)_Y^{\ell oc}$, are now under progress.

The predicted properties of the Z boson by the standard model are consistent with those observed by the precision measurements at the CERN Large Electron-Positron Collider, our predictions on the Z properties must at least reproduce those observed properties. The effects from $W'$ on the weak interaction phenomenology can be kept mild since it couples to the left-handed currents as $W$ does. In the present model of weak bosons, the standard model prediction will be recovered in the limit of $m_{W'} \to m_{Z'}$, yielding $s_{\alpha,\delta} \to 0$ even if $W'$ and $Z'$ are kept light. The simple computation leads to the following behavior of the coupling parameter, $\varepsilon$:

$$\varepsilon^2 \sin^2\theta \to \left((m_{Z'}/m_Z)^2 - 1\right)\left(1 - (m_{W'}/m_{Z'})^2\right). \tag{44}$$

The masses of $W'$ and $Z'$ are shown to satisfy the modified mass relation:

$$m_W m_{W'} = \cos\theta\, m_Z m_{Z'}, \tag{45}$$

which can be experimentally checked. The extensive studies on extra weak bosons have lately been made in Ref.[22] but our present case is not involved. To get definite conclusions about constraints on couplings and masses, we have to perform absolute evaluations including all the calculable radiative corrections [24], which will be presented elsewhere.

**Table Captions**

Table 1  Quantum numbers of quarks, leptons and Higgs scalar in $SU(3)_c^{loc} \times SU(2)_L^{loc} \times U(1)_Y^{loc} \times \boldsymbol{SU}(2)_L^{loc}$.

Table 2  Quantum numbers of subconstituents in $SU(3)_c^{loc} \times SU(2)_L^{loc} \times U(1)_Y^{loc}$.

**Figure Caption**

Figure 1 Possible physics of $W'$ and $Z'$ from "low"-energies to "high"-energies.

| | $SU(3)_c^{loc}$ | $SU(2)_L^{loc}$ | $\mathcal{SU}(2)_L^{loc}$ | $U(1)_Y^{loc}$ | $U(1)_{em}^{loc}$ |
|---|---|---|---|---|---|
| $q_L^A = \begin{pmatrix} u_L^A \\ d_L^A \end{pmatrix}$ | **3** | **1** | **2** | 2/3 | $\begin{pmatrix} 2/3 \\ -1/3 \end{pmatrix}$ |
| $\ell_L = \begin{pmatrix} \nu_{eL} \\ e_L \end{pmatrix}$ | **1** | **1** | **2** | -1 | $\begin{pmatrix} 0 \\ -1 \end{pmatrix}$ |
| $H$ | **1** | **1** | **2** | -1 | $\begin{pmatrix} 0 \\ -1 \end{pmatrix}$ |
| $\xi$ | **1** | **2** | **2** | 0 | $\begin{pmatrix} 0 & 1 \\ -1 & 0 \end{pmatrix}$ |

**Table 1** Quantum numbers of quarks and leptons

| | $SU(3)_c^{loc}$ | $SU(2)_L^{loc}$ | $U(1)_Y^{loc}$ | $U(1)_{em}^{loc}$ |
|---|---|---|---|---|
| $\begin{pmatrix} w_{1L} \\ w_{2L} \end{pmatrix}$ | **1** | **2** | 0 | $\begin{pmatrix} 1/2 \\ -1/2 \end{pmatrix}$ |
| $w_{1R}$ | 1 | 1 | 1 | 1/2 |
| $w_{2R}$ | 1 | 1 | -1 | -1/2 |
| $c^A$ | 3 | 1 | 1/3 | 1/6 |
| $c^0$ | 1 | 1 | -1 | -1/2 |

**Table 2** Quantum numbers of subconstituents

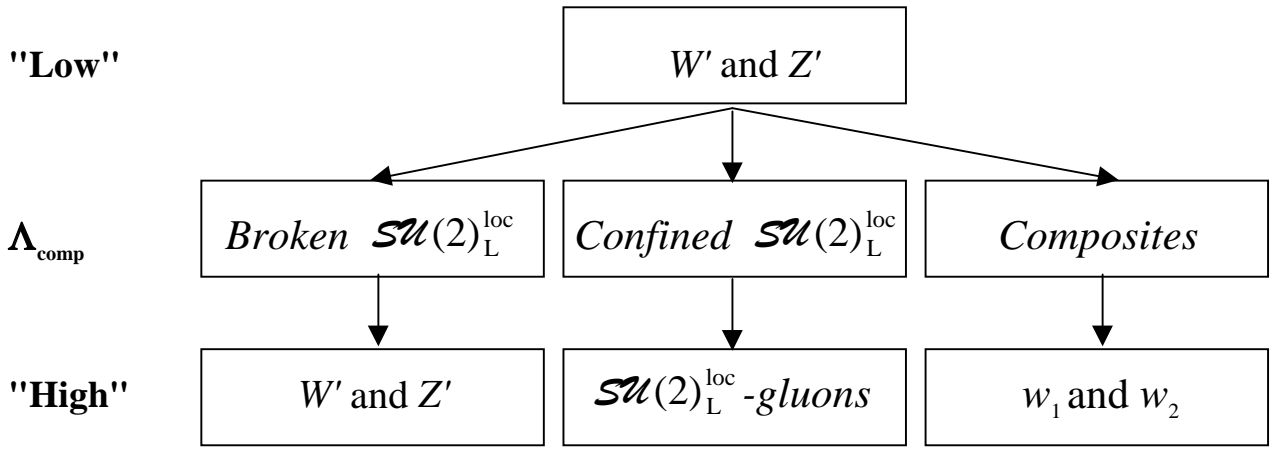

**Figure 1** Possible physics of $W'$ and $Z'$ from "low"-energies to "high"-energies